\documentclass[aps,prb,twocolumn,groupedaddress,floatfix,showpacs]{revtex4-1}

\usepackage{epsfig}
\usepackage{amssymb}
\usepackage{amsmath}
\usepackage{hyperref}

\begin{document}

\title{Renormalization group flow for fermions into antiferromagnetically ordered phases: Method and mean-field models}
\author{Stefan A. Maier and Carsten Honerkamp}
\affiliation{Institute for Theoretical Solid State Physics, RWTH Aachen University, D-52056 Aachen, Germany 
\\ and JARA - FIT Fundamentals of Future Information Technology}

\date{October 9, 2012}

\begin{abstract}
We present a functional renormalization group flow for many-fermion lattice models into phases with broken spin-rotational symmetry. The flow is expressed purely in terms of fermionic vertex functions. The symmetry breaking is seeded by a small initial anomalous self-energy which grows at the transition scale and which prevents a runaway-flow at nonzero scales. Focusing on the case of commensurate antiferromagnetism, we discuss how the interaction vertex can be parametrized efficiently. For reduced models with long-range bare interactions we show the results of standard mean-field theory are reproduced by the fRG and how anisotropies in the spin sector change the flows. We then describe a more efficient decomposition of the interaction vertex that should allow for the treatment of more general models.
\end{abstract}


\maketitle

 \section{Introduction}
Spontaneous symmetry breaking is an important building block of our understanding of interacting matter.  Typically, the symmetry breaking occurs at energy scales much smaller than those at which the corresponding theoretical models are formulated. This makes a quantitative theoretical description of these models and hence of the respective materials challenging. In a function renormalization group (fRG) approach (for a recent review, see Ref.~\onlinecite{metzner-rmp}), flow equations interpolate between the parameters of the microscopic (bare) action at a high energy scale and those of the one-particle irreducible functional, which can be regarded as an effective action at lower scales. 

For the description of spontaneous symmetry breaking in correlated electronic systems, one often decouples the two-particle interaction via a Hubbard-Stratonovich transformation. Then the fermions can be integrated out, and the remaining bosonic theory of the composite field(s) can possibly, as a first approximation, be treated in saddle-point approximation. However, in certain relevant cases such as unconventional superconductivity, the bare interaction does not contain the necessary attractive component in the desired channels, and a more suitable effective interaction needs to be constructed, e.g. by fermionic perturbation theory. Furthermore, even if this first problem is not present, the mean-field solution ignores potentially important corrections that can lead to changes in the resulting energy spectrum, or to the renormalization or even disappearance of the order due to collective fluctuations. Also, integrating out the fermions completely can lead to severe complications, e.g. due to diverging coupling constants of the higher order terms.\cite{loehneysen-rmp} In particular for the latter aspects, it may be advantageous to consider the renormalization-group flow for a mixed system consisting of fermions and exchange bosons. At the onset of symmetry breaking, the renormalized potential for the zero-momentum mode of such an exchange boson has degenerate minima corresponding to a nonzero order parameter.\cite{btw-review,metzner-rmp} In the symmetry-broken regime, the effective action may be expanded around one of these minima, giving rise to flow equations for the radial- and Goldstone-mode propagators and the fermionic gap, as in the case of a fermionic superfluid.\cite{strack-superfluid} In a series of publications,\cite{krahl_Hubbard_rebos,krahl_iAFM,friederich_10,friederich_11} the low-temperature physics of the two-dimensional (2D) Hubbard model, which may be regarded as a generic model for high-$ T_{\rm c} $ compounds, was analyzed within such an approach. In these papers, the generation of an attractive $d$-wave pairing interaction is taken into account by rebosonizing fermion-fermion interactions generated in the flow. In the partially bosonized formulation, however, the momentum dependence that is not captured by the bosonic channels has been dropped. Moreover, a Hubbard-Stratonovich decoupling of the bare interaction adds some bias to the description. For the Hubbard interaction, this problem has been circumvented by only bosonizing renormalizations of the interaction.\cite{friederich_11}

While these partially bosonized flow studies have their merits and should be extended further, one has to pay some more attention in order to capture effects (such as the generation of attractive pairing components) that one gets rather straightforwardly in purely fermionic approaches. In particular in cases with competing orders this might become more cumbersome. In such situations, purely fermionic flows have been used successfully at scales above the spontaneous symmetry breaking (for a recent review, see Ref.~\onlinecite{metzner-rmp}). 
So, in another string of works\cite{brokensy_SHML,gersch_cdw,gersch_firstorder,gersch_superfluid}, one has tried to continue these purely fermionic flows into the symmetry breaking regime. There is, however, a crucial difference to the mixed flows: If the bare action does not contain a symmetry-breaking seed field, the tendency toward spontaneous symmetry breaking in such a purely fermionic description manifests itself in a flow to strong coupling, such as a Cooper or spin-density-wave (SDW) instability. An instability analysis of that type has, among other systems, been performed for the 2D Hubbard model\cite{honerkamp_tflow,husemann_09,husemann_12} and the iron pnictides.\cite{pnictides} For a non-vanishing seed field,  mean-field models for broken translational\cite{gersch_cdw} and U(1) (Refs.~\onlinecite{brokensy_SHML,eberlein_param}) symmetry have been studied. These models are exactly solvable in the thermodynamic limit, as only diagrams in one channel contribute. In order to reproduce the exact result, the fermionic flow needs be considered in the Katanin truncation.\cite{katanin_trunc} Beyond mean-field models, superfluidity in the attractive 2D Hubbard model was studied within a purely fermionic description,\cite{gersch_superfluid} and recently in channel-decomposed approach.\cite{eberlein-unpub}

Phases of broken spin symmetry, in contrast, have not been studied yet with these extended RG flows. The competition  (and possible coexistence) of antiferromagnetism (AFM) and superconductivity (SC), however, is of prime interest in high-$ T_{\rm c} $ compounds and iron pnictides (see, for example, Refs.~\onlinecite{rohe,schmalian}).  Hence, fRG flows should also be continued into the SU(2)-broken phase.

The goal of this work is to lay the foundations for doing so in realistic models. It is organized as follows. In Sec.~\ref{sec:gen}, we give an efficient parametrization of a spin-symmetry-broken interaction and derive RG flow equations for the respective coupling functions. In the sequel, we analyze reduced mean-field models for a SDW phase. Section~\ref{sec:iso} is devoted to an isotropic interaction. The effective action is parameterized further and couplings corresponding to the bosonic radial and Goldstone modes are identified. The resulting flow equations are found to be equivalent to resummation within the random-phase approximation (RPA) and the self-consistent gap equation. The generalization to an anisotropic mean-field interaction in Sec.~\ref{sec:aniso} is straight forward. Before we finally conclude in Sec.~\ref{sec:conc}, we propose a channel decomposition of the flow equations in the SU(2)-broken phase in Sec.~\ref{sec:ch-dec}, which can also be used for an improved treatment of more general models.
 \section{General case} \label{sec:gen}

\subsection{Parametrization of the interaction} \label{sec:param-gen}
In Ref.~\onlinecite{salm_hon_2001}, the four-point vertex was parametrized for theories that remain SU(2)-symmetric in spin space. The solution of the flow equations for appropriate cases then yields a flow to strong coupling where at least one class of components of the four-point vertex diverges at a nonzero critical scale. If the Fermi surface is nested, e.g., due to flat parallel sides, and if the initial short-range interaction is repulsive, the flow to strong coupling usually indicates an antiferromagnetic instability where the static susceptibility in this channel becomes infinitely large. The interpretation of this instability is that below this critical scale, antiferromagnetic order sets in. This regularizes the growth of the four-point vertex in a reasonable way that is discussed further below, and definitely breaks the SU(2)-spin rotational symmetry. Hence, in order to understand the low-scale regime, one needs to continue the flow into the broken-symmetry regime. In this section we address the necessary parametrization of the interaction in the SU(2)-broken phase. In order to keep the resulting numerical calculations tractable, we will restrict ourselves to spin-$1/2$-fermions exhibiting collinear antiferromagnetism, i.e.\ we break the spin-symmetry only in one direction. More specifically, we allow for $ \langle S_z \rangle \neq 0 $ while we still require $ \langle S_x \rangle = \langle S_y \rangle = 0 $, where $ S_i $ denotes the $i$-th component of the spin operator. For such a case, a similar parametrization\cite{xxz_spinRG} has been given in real space.
The two-point Green's function only has diagonal entries in spin space and can then be split into a spin-flip symmetric and a spin-antisymmetric part
 $ G_{\tau_1,\tau_2} = G_1 \delta_{\tau_1,\tau_2} + G_z \sigma^z_{\tau_1,\tau_2} = G_{\tau_1}$, with  the Pauli $ z$-matrix $ \sigma^z $ and the spin indices  $\tau_i$ being $\uparrow$ or $\downarrow$. The elements of the remaining
spin symmetry group are $ U(\varphi)= e^{i \varphi \sigma^z} $ with arbitrary real $ \varphi $.  A general
charge conserving two-particle interaction can be written as an object quartic in fermionic Grassmann fields,
\begin{equation*}
 \frac{1}{4} \int \! d \xi_1 \, d \xi_2 \, d \xi_3 \, d \xi_4 \, f (\xi_1,\xi_2,\xi_3,\xi_4) \, \bar{\psi} (\xi_1) \bar{\psi} (\xi_2) \psi (\xi_3) \psi (\xi_4) \, , 
\end{equation*}
with generalized quantum numbers $\xi_i $, containing e.g. fermionic  Matsubara frequency, wave vector or momentum and spin quantum number along the $z$-direction, $\tau_i$. The integration $\int d\xi_i$ runs over all these entries.
Under the remaining symmetry transformations $U(\varphi)$, this term transforms according to
\begin{equation*}
 f (\xi_1,\xi_2,\xi_3,\xi_4) \to e^{i \varphi ( - \tau_1 - \tau_2 + \tau_3 + \tau_4)} f (\xi_1,\xi_2,\xi_3,\xi_4) \, ,
\end{equation*}
 implying conservation of the projected spin $z$-components, $ \tau_1 + \tau_2 = \tau_3 + \tau_4 $. There are six spin configurations satisfying this constraint, namely
\begin{equation*}
 \uparrow \uparrow \uparrow \uparrow \, , \quad
 \downarrow \downarrow \downarrow \downarrow \, , \quad
 \uparrow \downarrow \uparrow \downarrow \, , \quad
 \uparrow \downarrow \downarrow \uparrow \, , \quad
 \downarrow \uparrow \uparrow \downarrow \, , \quad
 \downarrow \uparrow \downarrow \uparrow \ .
\end{equation*}
Since  $f (\xi_1,\xi_2,\xi_3,\xi_4) $ is antisymmetric under $ \xi_1 \leftrightarrow \xi_2 $
and $ \xi_3 \leftrightarrow \xi_4 $,
the spin-dependence of the interaction can be parametrized using three independent momentum- and frequency-dependent (encoded in variables $k_i$) coupling functions
\begin{align} \notag
f (\xi_1,\xi_2,\xi_3,\xi_4) =&V_\uparrow (k_1,k_2,k_3,k_4) \, \delta_{{\boldsymbol \tau},\uparrow \uparrow \uparrow \uparrow} \\ \notag
 & + V_\downarrow (k_1,k_2,k_3,k_4) \, \delta_{{\boldsymbol \tau},\downarrow \downarrow \downarrow \downarrow} \\ \notag
& + V_{\uparrow\downarrow} (k_1,k_2,k_3,k_4) \, \delta_{{\boldsymbol \tau},\uparrow \downarrow \uparrow \downarrow}\\ \notag
 & - V_{\uparrow\downarrow} (k_1,k_2,k_4,k_3) \, \delta_{{\boldsymbol \tau},\uparrow \downarrow \downarrow \uparrow}\\ \notag
  &+ V_{\uparrow\downarrow} (k_2,k_1,k_4,k_3) \, \delta_{{\boldsymbol \tau},\downarrow \uparrow \downarrow \uparrow}\\ \label{eqn:param}
 & - V_{\uparrow\downarrow} (k_2,k_1,k_3,k_4) \, \delta_{{\boldsymbol \tau},\downarrow \uparrow \uparrow \downarrow} \, .
\end{align}
 Due to the antisymmetry property of $f (\xi_1,\xi_2,\xi_3,\xi_4) $, $ V_\uparrow $ and $ V_\downarrow $ are antisymmetric under $ k_1 \leftrightarrow k_2 $
and $ k_3 \leftrightarrow k_4 $, whereas $ V_{\uparrow\downarrow}$ does not obey a Pauli principle constraint. Particle-hole symmetry implies
\begin{align*}
 V_\uparrow (k_1,k_2,k_3,k_4) & = V_\uparrow (k_4,k_3,k_2,k_1)  \, , \\
 V_\downarrow (k_1,k_2,k_3,k_4) & = V_\downarrow (k_4,k_3,k_2,k_1)  \, , \\
V_{\uparrow\downarrow} (k_1,k_2,k_3,k_4) & = V_{\uparrow\downarrow} (k_3,k_4,k_1,k_2) \, ,
\end{align*}
while restoring the SU(2) symmetry imposes the constraint
\begin{align} \notag
 V_\uparrow & (k_1,k_2,k_3,k_4) = V_\downarrow (k_1,k_2,k_3,k_4) \\  \label{eqn:su2const}
& = V_{\uparrow\downarrow} (k_1,k_2,k_3,k_4) -V_{\uparrow\downarrow} (k_2,k_1,k_3,k_4) 
\end{align}
on the coupling functions.
A global SU(2) Ward identity can be derived as in the U(1) case in Ref.~\onlinecite{brokensy_SHML}, Eq.~(85). One obtains
\begin{align}  \notag
 C_z (k_1,k_2) -  C^0_z (k_1,k_2) & = 
 - \int \!\! d p_1 \,d p_2 \,d p_3 \,d p_4 \, C^0_z (p_1,p_2) \\ \notag
  & \quad \times G_\downarrow (p_2,p_3) \, G_\uparrow (p_4,p_1) \\ \label{eqn:genWI} 
 & \quad \times V_{\uparrow \downarrow} (k_1,p_3,p_4,k_2) \, ,
\end{align} 
where $ C_z $ and $ C^0_z $ denote the spin-antisymmetric part of the inverse of the full and the bare propagator, respectively. Note that $ G_\tau $  represents the full propagator and that $ V_{\uparrow \downarrow} $ enters as a renormalized interaction.

\subsection{Flow equations}
This new parametrization allowing for collinear spin order in the $z$ direction can now be inserted into the flow equations for the general one-particle-irreducible self-energy $ \Sigma (\xi_1,\xi_2) $ and four-point vertices $ f$ given in paragraph~4.1 of Ref.~\onlinecite{salm_hon_2001}, where only U(1) invariance (i.e., charge conservation) is assumed. These flow equations read as follows.
 The self-energy flows according to
\begin{equation} \label{eqn:SEorig}
 \partial_\lambda \Sigma (\xi_1,\xi_2) = \int \! d \eta_1 \, d \eta_2 \, S(\eta_2,\eta_1) \,
  f(\xi_1,\eta_1,\eta_2,\xi_2) \, ,
\end{equation}
where $ S $ denotes the single-scale propagator
\begin{align*}
  S  (\xi_1,\xi_2) & = \partial_\lambda G (\xi_1,\xi_2)
- \int \! d\eta_1 \, d\eta_2 \, G^{-1} (\xi_1,\eta_1) \\  & \quad \times \left[ \partial_\lambda \Sigma (\eta_1,\eta_2) \right] G^{-1} (\eta_2,\xi_2) \, .
\end{align*}
The flow of the interaction for a charge-conserving theory is given by
\begin{align*}
 \partial_\lambda  f (\xi_1,&\xi_2,\xi_3,\xi_4) = F_{\rm pp} (\xi_1,\xi_2,\xi_3,\xi_4) \\ &+
F_{\rm ph} (\xi_1,\xi_2,\xi_3,\xi_4) -
F_{\rm ph} (\xi_1,\xi_2,\xi_4,\xi_3) \, ,
\end{align*}
where
\begin{align*}
 F_{\rm pp} (\xi_1,\xi_2,\xi_3,\xi_4) & = \frac{1}{2} \int \!\! d \eta_1 \, d \eta_2 \, d \eta_3 \, d \eta_4 
\,  f(\xi_1,\xi_2,\eta_2,\eta_3) \,  \\ \times & 
 f (\eta_4,\eta_1,\xi_3,\xi_4) \left[ \partial_\lambda G(\eta_2,\eta_1) \, G(\eta_3,\eta_4) \right] \, ,
\end{align*}
\begin{align*}
 F_{\rm ph} (\xi_1,\xi_2,\xi_3,\xi_4) & = - \int \!\! d \eta_1 \, d \eta_2 \, d \eta_3 \, d \eta_4 \,
 f (\eta_4,\xi_2,\xi_3,\eta_1)  \\ \times &
 f (\xi_1,\eta_2,\eta_3,\xi_4)
 \left[ \partial_\lambda G(\eta_1,\eta_2) \, G(\eta_3,\eta_4) \right] \, .
\end{align*}
 In contrast to Ref.~\onlinecite{salm_hon_2001}, we have already performed a so-called Katanin substitution\cite{katanin_trunc} of the loops here, i.e.\ we have replaced single-scale propagators in the loops above by full scale-derivatives of the propagators in order to account for non-overlapping three-particle contributions.
 Note that this substitution does not affect the flow equation (\ref{eqn:SEorig}) for the self-energy.
 
Parametrizing the effective interaction according to Eq.~(\ref{eqn:param}) simplifies the numerical solution of the RG flow. The flow equations then read as follows. The right-hand side of the flow equations for the interaction 
 can be decomposed into particle-particle and particle-hole diagrams
 \begin{align} \notag
  \partial_\lambda V_\tau (k_1,k_2,k_3,k_4) & =  T^{\rm pp}_\tau (k_1,k_2,k_3,k_4)  \\
& \quad +  T^{\rm ph}_\tau (k_1,k_2,k_3,k_4)  \, , \label{eqn:vup}
 \end{align}
 \begin{align} \notag
  \partial_\lambda V_{\uparrow\downarrow} (k_1,k_2,k_3,k_4) & =  T^{\rm pp}_{\uparrow\downarrow} (k_1,k_2,k_3,k_4) \\ & \quad +  T^{\rm ph}_{\uparrow\downarrow} (k_1,k_2,k_3,k_4)  \, . \label{eqn:vupdown} 
 \end{align}
 With the short-hand notation $ L_{\tau_1,\tau_2} (p,p';q,q') = \partial_\lambda \left[ G_{\tau_1} (p,p') \, G_{\tau_2} (q,q') \right] $
 for the loop diagrams, we obtain the following contributions:
\begin{widetext}
 \begin{equation} \label{eqn:start-ffe}
T^{\rm pp}_\uparrow (k_1,k_2,k_3,k_4) = \frac{1}{2} \int\! dp \, d p' \, dq \, dq' \, V_\uparrow (k_1,k_2,p,q) \, V_\uparrow (q',p',k_3,k_4) \, L_{\uparrow,\uparrow} (p,p';q,q') \, ,
 \end{equation}
\begin{align} \notag
T^{\rm ph}_\uparrow (k_1,k_2,k_3,k_4) =&- \! \int\! dp \, d p' \, dq \, dq' \,
 \left[ V_\uparrow (q',k_2,k_3,p) \, V_\uparrow (k_1,p',q,k_4) - V_\uparrow (q',k_2,k_4,p) \, V_\uparrow (k_1,p',q,k_3) \right] \, L_{\uparrow,\uparrow} (p,p';q,q') \\ \notag
&- \! \int\! dp \, d p' \, dq \, dq' \,
 \left[ V_{\uparrow\downarrow} (k_2,q',k_3,p) \, V_{\uparrow\downarrow} (k_1,p',k_4,q) -V_{\uparrow\downarrow} (k_2,q',k_4,p) \, V_{\uparrow\downarrow} (k_1,p',k_3,q)
 \right] \\ & \quad \times \, L_{\downarrow,\downarrow} (p,p';q,q') \, .
 \end{align}
 Likewise, we have
 \begin{equation}
T^{\rm pp}_\downarrow (k_1,k_2,k_3,k_4) = \frac{1}{2} \int\! dp \, d p' \, dq \, dq' \, V_\downarrow (k_1,k_2,p,q) \, V_\downarrow (q',p',k_3,k_4) \, L_{\downarrow,\downarrow} (p,p';q,q') \, ,
 \end{equation}
 \begin{align} \notag
T^{\rm ph}_\downarrow (k_1,k_2,k_3,k_4) =&- \! \int\! dp \, d p' \, dq \, dq' \,
 \left[ V_\downarrow (q',k_2,k_3,p) \, V_\downarrow (k_1,p',q,k_4) - V_\downarrow (q',k_2,k_4,p) \, V_\downarrow (k_1,p',q,k_3) \right] \, L_{\downarrow,\downarrow} (p,p';q,q') \\ \notag
&- \! \int\! dp \, d p' \, dq \, dq' \,
 \left[ V_{\uparrow\downarrow} (q',k_2,p,k_3) \, V_{\uparrow\downarrow} (p',k_1,q,k_4) -V_{\uparrow\downarrow} (q',k_2,p,k_4) \, V_{\uparrow\downarrow} (p',k_1,q,k_3)
\right] \\ & \quad \times \, L_{\uparrow,\uparrow} (p,p';q,q') 
 \end{align}
and
 \begin{align}
T^{\rm pp}_{\uparrow\downarrow} (k_1,k_2,k_3,k_4) =& - \! \int\! dp \, d p' \, dq \, dq' \, V_{\uparrow\downarrow} (k_1,k_2,p,q) \, V_{\uparrow\downarrow} (p',q',k_3,k_4) \, L_{\uparrow,\downarrow} (p,p';q,q') \, ,\\ \notag
T^{\rm ph}_{\uparrow\downarrow} (k_1,k_2,k_3,k_4) =&- \! \int\! dp \, d p' \, dq \, dq' \,
 V_{\uparrow\downarrow} (q',k_2,k_3,p) \, V_{\uparrow\downarrow} (k_1,p',q,k_4) \, L_{\downarrow,\uparrow} (p,p';q,q') \\ \notag
&- \! \int\! dp \, d p' \, dq \, dq' \,
 V_{\uparrow\downarrow} (q',k_2,p,k_4) \, V_\uparrow (k_1,p',q,k_3) \, L_{\uparrow,\uparrow} (p,p';q,q') \\ \label{eqn:stop-ffe}
&- \! \int\! dp \, d p' \, dq \, dq' \,
 V_\downarrow (q',k_2,k_4,p) \, V_{\uparrow\downarrow} (k_1,p',k_3,q) \,  L_{\downarrow,\downarrow} (p,p';q,q') \, .
 \end{align}
For the flow of the self-energy, we find
\begin{align}
 \partial_\lambda \Sigma_\uparrow (k_1,k_2) & = - \int \! d p \, d p' \left[ S_\uparrow (p,p') \, V_\uparrow (k_1,p',p,k_2) - S_\downarrow (p,p') \, V_{\uparrow\downarrow} (k_1,p',k_2,p) \right] \label{eqn:sigmaup} \\
 \partial_\lambda \Sigma_\downarrow (k_1,k_2) & = -\int \! d p \, d p' \left[ S_\downarrow (p,p') \, V_\downarrow (k_1,p',p,k_2) - S_\uparrow (p,p') \, V_{\uparrow\downarrow} (p',k_1,p,k_2) \right] \, . \label{eqn:sigmadown}
\end{align}
\end{widetext}
The first-order differential equations (\ref{eqn:vup}), (\ref{eqn:vupdown}), (\ref{eqn:sigmaup}), and (\ref{eqn:sigmadown}) have to be solved together. We have not specified the flow parameter $\lambda$ yet, and these equations hold for any continuous parameter that appears in the quadratic part of the initial action only. In the following, we will use a sharp band-energy cutoff, where the modes below scale $\lambda$ are suppressed in the free propagator.
\section{Mean-field model for commensurate antiferromagnetism} \label{sec:iso}

\subsection{Parametrization}
Before we discuss more complicated cases, we study a simple mean field model for fermions on a $D$-dimensional lattice with infinitely long-ranged staggered (i.e.\ antiferromagnetic) spin-spin interactions.
More precisely, we consider lattice fermions on a $ D $-dimensional torus with circumference $ L $ in all directions and take the limit $ L \to \infty $. The staggering is characterized by a wave vector ${\bf Q}$, which does not need to be specified in full detail. However, for simplicity, we restrict our analysis to commensurate ordering vectors $ {\bf Q} $. 
The action then reads as
\begin{equation} \label{eqn:MFmodel}
 S= \int\!dk \, \bar{\psi}_\tau (k) \, \left( i k_0 - \epsilon_{\bf k} \right) \psi_\tau (k) + \frac{J}{\Omega} \,  {\bf S}_Q\cdot {\bf S}_{-Q} \, ,
\end{equation}
where $ {\bf S}_Q = \int\!dk \, \bar{\boldsymbol \psi} (k) \, {\boldsymbol \sigma}\, {\boldsymbol \psi} (k+Q) $ with $ Q= (0,{\bf Q}) $ and $ J > 0 $. 
Later, we will also assume perfect particle-hole nesting entailed by a fermionic lattice dispersion with the property $ \epsilon_{\bf k} = - \epsilon_{{\bf k}+{\bf Q}} $.  This assumption is not crucial for the validity of the scheme, but makes the solution of the flow equations much simpler.
 Note that the interaction term in Eq.~(\ref{eqn:MFmodel}) contains in total only two $k$-summations, i.e., the frequency- and momentum structure of the interaction is strongly or doubly restricted compared to the general translationally invariant case with three summations for momenta and frequencies.
Its strength is renormalized by the $ 1+D $-dimensional volume $ \Omega = \beta L^D$ at temperature $ 1 /\beta $. In the following, we will consider the RG flow in the limit $ \Omega \to \infty $, which corresponds to the thermodynamic limit and/or zero temperature.

In order to break both SU(2) and translational invariance, we add a spin-antisymmetric term $\Delta  {\bf S}^z_Q $
to the quadratic part of the bare action.
In our parametrization the quartic part of Eq.~(\ref{eqn:MFmodel}) corresponds to the following coupling functions in the ultraviolet:
\begin{align*}
 V_\uparrow^\infty &= \frac{2 J}{\Omega} \left[ \delta \left( k_2 - k_3 + Q \right) - \delta \left( k_1 - k_3 + Q \right) \right] \delta_{ \{ k_i \} } \, , \\
 V_\downarrow^\infty &= \frac{2 J}{\Omega} \left[ \delta \left( k_2 - k_3 + Q \right) - \delta \left( k_1 - k_3 + Q \right) \right] \delta_{\{k_i\}} \, , \\
 V_{\uparrow\downarrow}^\infty &= \frac{2 J}{\Omega} \left[ 2 \delta \left( k_2 - k_3 + Q \right) + \delta \left( k_1 - k_3 + Q \right) \right] \delta_{\{k_i\}} \, .
\end{align*}
Here, the superscript denotes the scale and we have introduced $ \delta ( p-q) = \Omega \,  \delta_{p_0,q_0} \delta_{{\bf p},{ \bf q}} $ and
\begin{equation*}
  \delta_{\{k_i\}} = \delta \left( k_1 + k_2 - k_3 - k_4 \right)
\end{equation*} 
as a short-hand notation for an energy- and momentum-conserving $ \delta $ function.
 
 Here and throughout, we use a pseudo-continuous notation, i.e., we interpret momentum and frequency integrals $ \int \! d  k $ as a summation $ \Omega^{-1} \sum_{k} $ 
over the discrete momenta of the finite system  at finite temperature
in the thermodynamic limit $ L \to \infty  $ and send $ \beta \to \infty $ for zero-temperature calculations. As pointed out in Refs.~\onlinecite{brokensy_SHML,gersch_cdw,eberlein_param}, only diagrams in one channel contribute in reduced mean-field theories. In the case considered here, this implies that particle-particle diagrams and diagrams with overlapping loops vanish in the thermodynamic limit since the square of a $ \delta $ function brings in an extra factor $\Omega$ in non-overlapping particle-hole loops. Particle-hole ladders are therefore of first order in $ \Omega^{-1} $ just as the bare interaction. In the thermodynamic limit, only these first-order contributions to the
 renormalized interaction survive and hence a Bethe-Salpether equation
\begin{align*}
 f^\lambda &(\xi_1,\xi_2,\xi_3,\xi_4)=- {\cal A}_{\xi_3,\xi_4} \int\! d\eta_1 \cdot \cdot \, d \eta_4 \, G^\lambda (\eta_1,\eta_2) \\
 & \times G^\lambda (\eta_3,\eta_4) \, f^\infty (\eta_4,\xi_2,\xi_3,\eta_1) \, f^\lambda (\xi_1,\eta_2,\eta_3,\xi_4) 
\end{align*}
holds, where the operator $ {\cal A} $ antisymmetrizes the coupling function according to $ {\cal A}_{\xi_1,\xi_2} h (\xi_1,\xi_2) = \left[ h \left(\xi_1,\xi_2 \right) - h \left( \xi_2, \xi_1 \right) \right] / 2 $. All diagrams not resummed by this equation are at least of second order in $ \Omega^{-1} $. Note that the Bethe-Salpether equation holds without any further assumption for the dispersion such as perfect nesting.
 Similarly, only contributions to the self-energy of zeroth order in $ \Omega^{-1} $ are retained in the thermodynamic limit.
 
For broken translational invariance, contributions to the interaction
 that violate momentum conservation are generated
during the flow in general. In the thermodynamic limit, however, such interaction terms vanish due to the restricted momentum dependence of the bare interaction. The renormalized
interaction can therefore be parametrized by four couplings
\begin{align} \notag
 V_\uparrow^\lambda &=  J_\uparrow^\lambda \left[ \delta \left( k_2 - k_3 + Q \right) - \delta \left( k_1 - k_3 + Q \right) \right] \delta_{ \{ k_i \} } / \Omega \, , \\ \notag
 V_\downarrow^\lambda &=  J_\downarrow^\lambda \left[ \delta \left( k_2 - k_3 + Q \right) - \delta \left( k_1 - k_3 + Q \right) \right] \delta_{ \{ k_i \} }  / \Omega \, , \\ \label{eqn:paramint}
 V_{\uparrow\downarrow}^\lambda &= \left[ J_{xy}^\lambda \delta \left( k_2 - k_3 + Q \right) + J_z^\lambda \delta \left( k_1 - k_3 + Q \right) \right] \delta_{\{k_i\}}  / \Omega \, .
\end{align}
 In the SU(2) symmetric case, these couplings fulfill the constraints $ J_\uparrow = J_\downarrow= J_{xy} - J_z $ [cf.\ Eq.~(\ref{eqn:su2const})]. In the absence of a density-density term in the interaction, we further have $ J_{xy} =2 J_z $.  
\subsection{Flow equations} \label{sec:mf-fleq}
  From the ultraviolet values $ J^\infty_\uparrow = J^\infty_\downarrow = 2 J $, $ J_{xy}^\infty = 4 J $ and $ J_z^\infty = 2 J $ these couplings flow according to
\begin{align}\label{eqn:startfleq}
 \dot{J}_\uparrow &= - J_\uparrow^2 \dot{B}_{\uparrow,\uparrow}  - J_z^2 \dot{B}_{\downarrow,\downarrow} \, , \\\label{eqn:fleq-alld}
 \dot{J}_\downarrow &= - J_\downarrow^2 \dot{B}_{\downarrow,\downarrow}  - J_z^2 \dot{B}_{\uparrow,\uparrow} \, , 
\end{align}
\begin{align}
 \dot{J}_z &= - J_z \left( J_\uparrow \dot{B}_{\uparrow,\uparrow} + J_\downarrow  \dot{B}_{\downarrow,\downarrow} \right) \, , \\
 \dot{J}_{xy} &= - J_{xy}^2 \dot{B}_{\downarrow,\uparrow} \, ,
\end{align}
with the dots denoting scale derivatives.
 For convenience, we have now suppressed the scale dependence in the notation. 
The particle-hole bubble
\begin{align*}
 {B}_{\tau_1,\tau_2} = \frac{1}{\Omega} \int \! dk \, dk' & \left[  G_{\tau_1} (k,k') \, G_{\tau_2} (k'+Q,k+Q)  \right.\\
& \left. + \, G_{\tau_1} (k,k'+Q) \, G_{\tau_2} (k'+Q,k) \right]
\end{align*}
corresponds to a trace in Nambu space with spinors
\begin{equation}  \label{eqn:Nambu-spinor}
{\boldsymbol \Psi}_\tau (k) = \left( \begin{array}{l} \psi_\tau (k) \\ \psi_\tau (k+Q) \end{array} \right) \, .
\end{equation}
 
The restricted momentum dependence of the interaction
leaves the momentum-conserving component of the self-energy unchanged during the flow. For the remaining anomalous%
\footnote{Throughout this paper, the adjective 'anomalous' refers to the momentum non-conserving components of the one-particle propagator (off-diagonal components in Nambu representation Eq.~(\ref{eqn:Nambu-spinor})).}
 components, the flow equations read as
\begin{align} \notag
\Sigma_\tau (k,k')  &= \Sigma_\tau \, \delta (k -k'+Q) \, , \\
 \dot{\Sigma}_\uparrow &= - J_\uparrow A_\uparrow + J_z A_\downarrow \, , \\ \label{eqn:stopfleq}
 \dot{\Sigma}_\downarrow &= - J_\downarrow A_\downarrow + J_z A_\uparrow \, ,
\end{align}
 with 
 \begin{equation*}
 A_\tau = \frac{1}{\Omega} \int\! dk\, dk' \, S_\tau (k,k') \, \delta (k-k'+Q)
 \end{equation*}
being the anomalous tadpole bubble.
 Note that if we had only broken the translational, but not the spin symmetry,
	 the loop integrals $ A_\tau $ and $ \dot{B}_{\tau_1,\tau_2} $ in the flow equations~(\ref{eqn:startfleq})--(\ref{eqn:stopfleq}) would not depend on the spin indices and the number of independent couplings would be reduced to one according to $ J_\uparrow = J_\downarrow= J_{xy} - J_z = J_z $, since the bare interaction does not contain a density-density term. 

If the symmetry breaking term in the bare action is antisymmetric under a spin flip as in our case, the property $ -\Sigma_\downarrow =  \Sigma_\uparrow =: \Delta $ is preserved in the flow.
We thus have $ {B}_{\uparrow,\uparrow} = {B}_{\downarrow,\downarrow} $, $ A_\uparrow = - A_\downarrow $, and $ J_\uparrow = J_\downarrow = J_z $ 
and thus the flow equations~(\ref{eqn:startfleq}), (\ref{eqn:fleq-alld}) and~(\ref{eqn:stopfleq}) become redundant. This way, the interaction
now remains invariant under spin flips. The flow equations to be solved finally read as
\begin{align} \label{eqn:flow-13}
 \dot{J}_z &= - 2 J_z^2 \dot{B}_{\uparrow,\uparrow} \, , \\ \label{eqn:flow-23}
 \dot{J}_{xy} &= - J_{xy}^2 \dot{B}_{\downarrow,\uparrow} \, , \\ \label{eqn:flow-delta}
 \dot{\Delta} & = - 2 J_z A_\uparrow \, .
\end{align}
Note that the flow of the in-plane coupling $ J_{xy} $ does not feed back on the flow of $ \Delta $ and $ J_z $. This behavior is similar to the reduced BCS model,\cite{brokensy_SHML} where the flow of the amplitude vertex and the gap decouples from the flow of the Goldstone vertex as well. Indeed, $ J_{xy} $ will turn out to account for the Goldstone modes further below.

So far, we have not made any specific assumption for the dispersion.
In the following, we will only consider the case of perfect nesting $ \epsilon_{\bf k} = - \epsilon_{{\bf k}+{\bf Q}} $ in order to simplify the integration of the loops.%
\footnote{If this perfect-nesting condition is violated by a chemical potential, a behavior similar to Ref.~\onlinecite{gersch_firstorder} is to be expected. A potential first-order phase transition may then be assessed by introducing a counterterm in the RG flow.}
Since the anomalous self-energy $ \Sigma_\tau $ is independent of $ k $, we may write in Nambu representation
\begin{align*}
 {\boldsymbol G}_\tau (k,k') = {\boldsymbol G}_\tau (k) \, & \delta (k-k') \, , \\
 {\boldsymbol G}_\tau (k) = \frac{\chi}{k_0^2 + \epsilon_{\bf k}^2 + \chi^2 \Sigma_\tau^2} & \left( \begin{array}{cc} 
-i k_0 -  \epsilon_{\bf k} & - \chi \Sigma_\tau \\ 
- \chi \Sigma_\tau & -i k_0 +  \epsilon_{\bf k} \end{array} \right) \, ,
\end{align*}
with a sharp energy-shell regulator $ \chi = \Theta ( |\epsilon_{\bf k} | -\lambda ) $.
 The expressions for the loops are now evaluated as follows.
 The Matsubara sums are performed analytically and Morris's lemma\cite{morris_lemma} is used to evaluate Heaviside functions at their jumps. For the sake of a lean notation, we define
\begin{align*}
 E & = \sqrt{ \epsilon^2 + \chi^2 \Delta^2} \, ,\\
 t & = \tanh \left( \frac{E}{2 T} \right) \, , \\
 t' & = \left[ 2T \cosh \left( \frac{E}{2 T} \right)^2 \right]^{-1} \, ,
\end{align*}
where $ T $ denotes the temperature. Calculating the Matsubara sum analytically, we obtain for the particle-hole bubble with opposite spins
 \begin{align*}
  \dot{B}_{\downarrow,\uparrow} & = - \partial_\lambda \int \! d {\bf k} \, \chi \, \frac{t}{2E}  \\
=    \left. \rho_\lambda \frac{ t}{2 E} \right|_{\epsilon=\lambda,\chi=1} \!\! & + \dot{\Delta} \Delta \int_\lambda^W \! d \epsilon \, \rho_\epsilon \left[ \frac{t}{E^3} - \frac{t'}{E^2} \right] \, ,
 \end{align*}
whereas the equal-spin bubble reads as
 \begin{align*}
 \dot{B}_{\uparrow,\uparrow}  &= \dot{B}_{\uparrow,\downarrow} + \partial_\lambda \int \! d{\bf k} \, \chi \, \Delta^2 \left( \frac{t}{2 E^3} - \frac{t'}{2 E^2} \right) \\
&= \rho_\lambda \left( \frac{t \, \lambda^2}{2 E^3} + \frac{\Delta^2 \, t'}{2 E^2} \right)_{\epsilon=\lambda,\chi=1} \\
 + & \dot{\Delta} \Delta \int_\lambda^W \! d \epsilon \, \rho_\epsilon  \left[ 3 \left( \frac{t}{E^3} - \frac{t'}{E^2} \right) \frac{\epsilon^2}{E^2} + \frac{t \, t' \Delta^2}{T E^3}
\right] \, .
 \end{align*}
Here, momentum integrals have been replaced by energy integrals with the density of states $ \rho_\epsilon $ and ultraviolet cutoff $ W $. For the tadpole bubble, Morris's lemma yields
\begin{equation*}
 A_\uparrow =  \rho_\lambda \left. \Delta \frac{t}{2 E} \right|_{\epsilon=\lambda,\chi=1}
\, .
\end{equation*}
If spin symmetry was restored, we would have to encounter a flow to strong coupling, since then $ \dot{\Delta} =0 $ and hence $ \dot{B}_{\tau_1,\tau_2} > 0 $.

\begin{figure}
 \includegraphics[width=8.5cm]{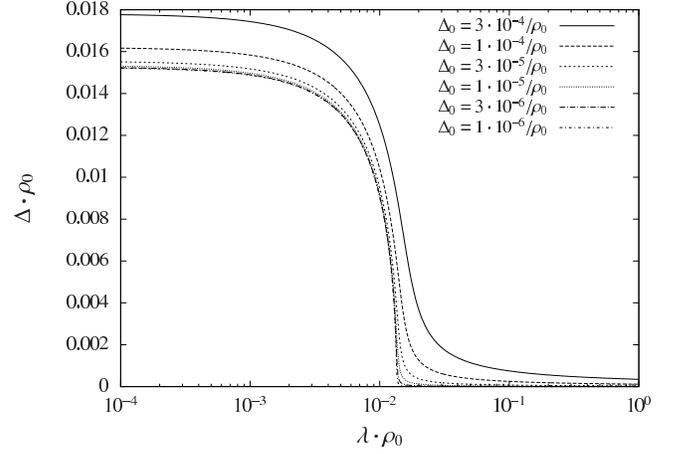}
 \caption{Flow of the gap $\Delta $ for zero temperature, perfect nesting, and a constant density of states $\rho_0$
 ($ W= 2 / \rho_0 $, $ J = 0.1 / \rho_0$).} \label{fig:delta}
\end{figure}
\begin{figure}
 \includegraphics[width=8.5cm]{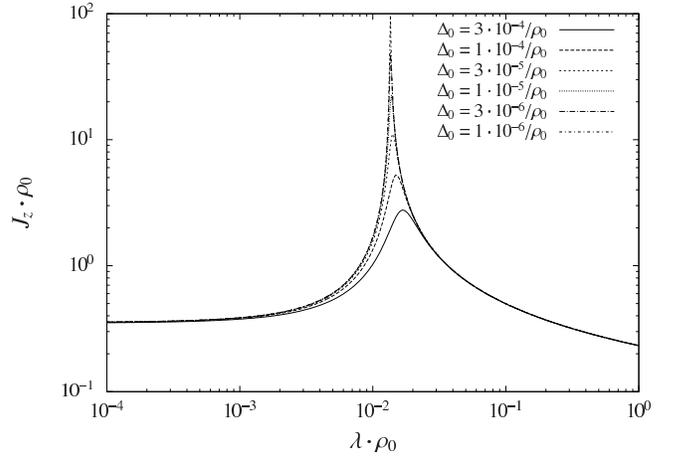}
 \caption{Flow of $J_z $ for zero temperature, perfect nesting, and a constant density of states $\rho_0$
 ($ W= 2 /\rho_0 $, $ J = 0.1 /\rho_0$).} \label{fig:j13}
\end{figure}
\begin{figure}
 \includegraphics[width=8.5cm]{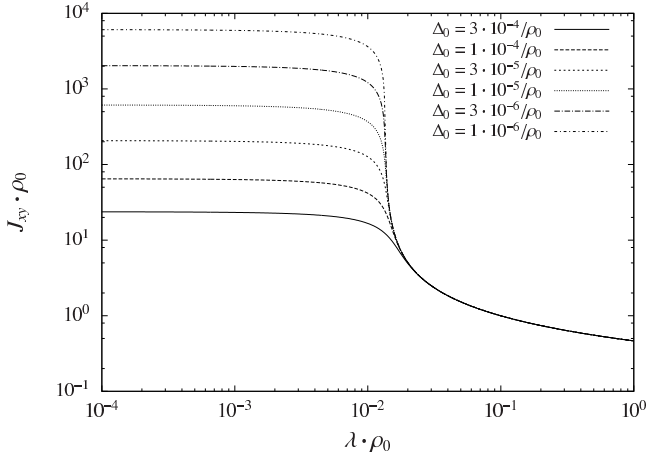}
 \caption{Flow of $J_{xy} $ according to Eqs.~(\ref{eqn:flow-13}), (\ref{eqn:flow-delta}), and~(\ref{eqn:ex-j23}) for zero temperature, perfect nesting and a constant density of states $\rho_0$
 ($ W= 2 /\rho_0 $, $ J = 0.1 /\rho_0$). Instead of the flow equation (\ref{eqn:flow-23}) for $ J_{xy} $, the formal solution (\ref{eqn:ex-j23}) has been used in order to avoid numerical complications.} \label{fig:j23}
\end{figure}
In Figs.~\ref{fig:delta}--\ref{fig:j23}, the flow of $ \Delta$, $ J_z $, and $J_{xy} $ is depicted for perfect nesting, zero temperature and a constant density of states $ \rho_0 $. At a critical scale $ \lambda_{\rm crit} \propto e^{- 1/(4J \rho_0)} $ (about $ 10^{-2} / \rho_0 $ in the case considered here), the coupling constants of the interaction grow and the gap opens.
At lower scales, $ J_z $ is then depressed by this gap while $ J_{xy} $ continues to grow.  The growth of $ J_z $ slightly above the critical scale occurs as a remainder of the AFM instability encountered in an SU(2)-symmetric flow.
If $ \Delta_0$ is sent to zero, the infrared values of $ J_z $ and $ \Delta $ saturate, whereas
 $ J_{xy} $ grows without bound in the infrared.
\subsection{Exact solution}
The flow equations (\ref{eqn:flow-13}) and (\ref{eqn:flow-23}) can be formally solved  to reproduce the RPA result
\begin{align} \label{eqn:ex-13}
 J_z &= \frac{2 J}{1 + 4 J \, {B}_{\uparrow,\uparrow} } \, ,\\
 J_{xy} &= \frac{4 J}{1 + 4 J \, {B}_{\uparrow,\downarrow} } \, .
\end{align}
The gap equation for the anomalous part of the propagator can now be derived if the tadpole is recast as
\begin{equation*}
 A_\uparrow = - \dot{\Delta} \, B_{\uparrow,\uparrow} + \frac{1}{\Omega} \int \!\! d k \, dk' \, \dot{G}_\uparrow (k,k') \, \delta (k-k'+Q) \,  .
\end{equation*}
Inserting Eq.~(\ref{eqn:ex-13}) into (\ref{eqn:flow-delta}) and integrating then yields the following gap equation:
\begin{align} \notag
 \Delta - \Delta_0 & = - \frac{4 J}{\Omega} \, \int \!\! d k \, dk' \, G_\uparrow (k,k') \, \delta (k-k'+Q) \\ \label{eqn:gap}
  & = - 4 J \, B_{\uparrow,\downarrow} \, \Delta \, ,
\end{align}
which can as well be obtained from a self-consistent mean-field ansatz.
For zero temperature, perfect nesting and a constant density of states, the gap behaves as $ \Delta \propto e^{-1/(4J \rho_0)} $. Since we now have 
\begin{equation} \label{eqn:ex-j23}
 J_{xy} = 4 J \, \Delta/\Delta_0 \, ,
\end{equation}
 $ J_{xy} $ diverges for $ \Delta_0, \lambda \to 0 $ and can thus 
be interpreted as a Goldstone vertex. $ J_z $, however, remains finite in the infrared and therefore corresponds to
 radial fluctuations of the staggered magnetization. The divergence of $ J_{xy} $ for vanishing $ \Delta_0 $ also directly
 follows from the Ward identity Eq.~(\ref{eqn:genWI}), which simplifies to
\begin{equation*}
 \Delta - \Delta_0 = - \Delta_0 \, J_{xy}\, B_{\uparrow,\downarrow} \, .
\end{equation*}

\section{Generalized, anisotropic mean-field model} \label{sec:aniso}
 In this section, we consider an interaction term of $xxz$-type 
\begin{align} \notag
 S = \int\!dk &\, \bar{\psi}_\tau (k) \, \left( i k_0 - \epsilon_{\bf k} \right) \psi_\tau (k) + \frac{J}{\Omega}\,  {\bf S}_Q\cdot {\bf S}_{-Q} \\ \label{eqn:gen_model}
 & + \frac{\alpha J}{\Omega} \, S_Q^z S_{-Q}^z \, ,
\end{align}
 which breaks the SU(2)-symmetry for a non-vanishing anisotropy $ \alpha $. Equation~(\ref{eqn:gen_model}) interpolates between the isotropic case  and an Ising-type interaction at $ \alpha \to \infty $ for positive $ \alpha $ or an interaction of $ xy $-type for $ -1 < \alpha < 0 $. In the following, the former will be referred to as the Ising regime and the latter as the $ xy $-regime. The RG flow equations can be parametrized in exactly the same way as for the mean-field model in Eq.~(\ref{eqn:MFmodel}). Again, this leads to the flow equations~(\ref{eqn:flow-13})--(\ref{eqn:flow-delta}), but
 now with the initial conditions $ J_z = 2 J (1 + \alpha) $, $ J_{xy} = 4 J $, and $ \Delta = \Delta_0 $ in the ultraviolet. 
The formal solution of the flow equations now reads as
\begin{align} \label{eqn:xxz13}
 J_z &= \frac{2 J\left(1+ \alpha \right) }{1 + 4 J \left(1+ \alpha \right) {B}_{\uparrow,\uparrow} } \, ,\\
\label{eqn:xxz23} J_{xy} &= \frac{4 J}{1 + 4 J \, {B}_{\uparrow,\downarrow} } \, , \\ \notag
& \\
\label{eqn:xxzdelta} \Delta - \Delta_0 & = - 4 J \left(1+ \alpha \right) \, B_{\uparrow,\downarrow} \, \Delta \, .
\end{align}
 The gap equation (\ref{eqn:xxzdelta}) can be derived analogously to Eq.~(\ref{eqn:gap}).
 Note that the Ward identity Eq.~(\ref{eqn:genWI}) has been derived for a spin-symmetric interaction and thus no longer holds.  So, a constraint that forces $ J_{xy} $ to diverge in the limit $ \Delta_0 \to 0 $ is now absent: for non-zero positive $ \alpha $, the Goldstone modes are gapped. As a consequence, the denominator of Eq.~(\ref{eqn:xxz23}) no more needs to be zero for $ \Delta_0, \lambda \to 0 $, but may now take on finite values. 
 
\begin{figure}
 \includegraphics[width=8.5cm]{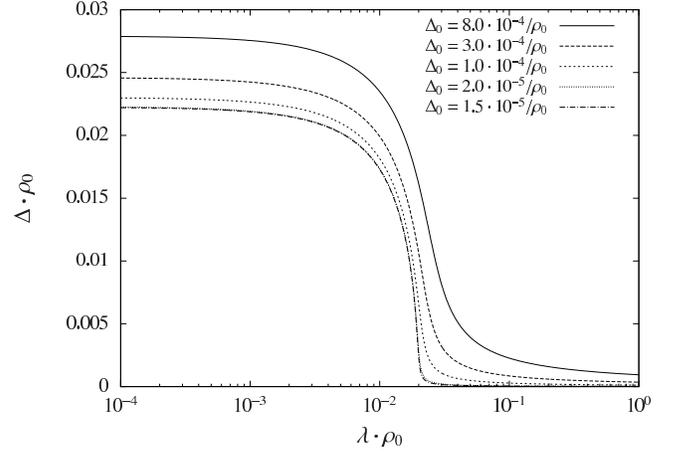}
 \caption{Flow of the gap $\Delta $ for $\alpha=0.2$, zero temperature, perfect nesting, and a constant density of states $\rho_0$
 ($ W= 2 / \rho_0 $, $ J = 0.09 / \rho_0$).} \label{fig:delta-alpha}
\end{figure}
\begin{figure}
 \includegraphics[width=8.5cm]{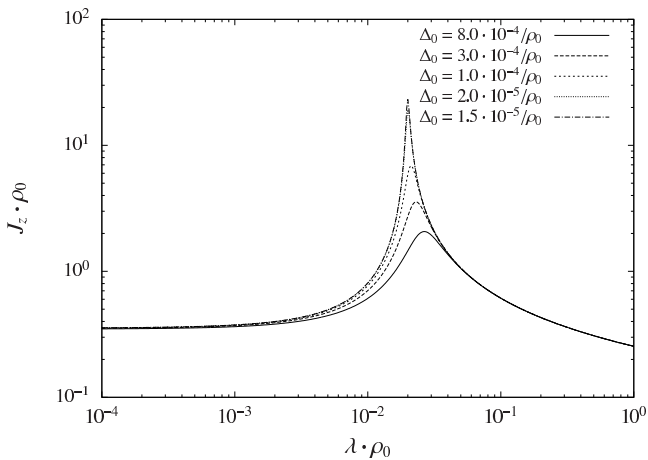}
 \caption{Flow of $J_z $ for $ \alpha=0.2$, zero temperature, perfect nesting, and a constant density of states $\rho_0$
 (all other parameters as in Fig.~\ref{fig:delta-alpha}).} \label{fig:j13-alpha}
\end{figure}
\begin{figure}
 \includegraphics[width=8.5cm]{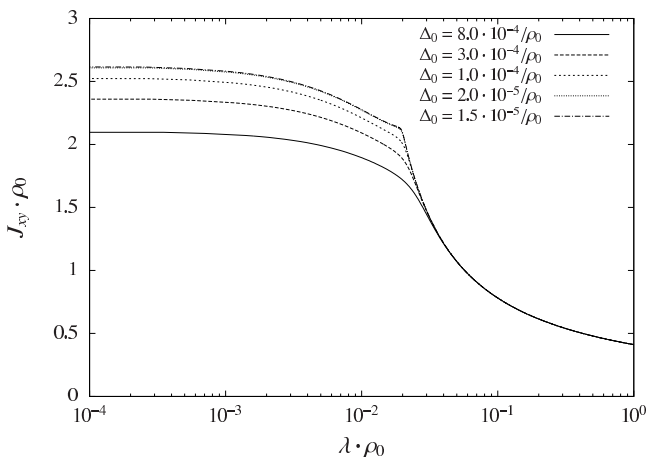}
 \caption{Flow of $J_{xy} $ for $ \alpha = 0.2$, zero temperature, perfect nesting, and a constant density of states $\rho_0$
 (all other parameters as in Fig.~\ref{fig:delta-alpha}). The curves for different values of the bare gap $ \Delta_0 $ become discernible below the critical scale.} \label{fig:j23-alpha}
\end{figure}
We now discuss the case of relatively weak anisotropies ($ \alpha = \pm 0.2$) for $ J >0 $ both in the Ising and in the $ xy$-regimes. In Figs.~\ref{fig:delta-alpha}--\ref{fig:j23-alpha}, the flow of the three couplings is depicted for $ \alpha=+0.2 $.
The flow of $ J_z $ in Fig.~\ref{fig:j13-alpha} again is peaked at some critical scale $ \lambda_{\rm crit} $, which is now enhanced as can be seen from Eq.~(\ref{eqn:xxz13}). At this scale, the gap in Fig.~\ref{fig:delta-alpha} starts to grow significantly and reaches a saturation
value in the infrared. 

So far the results only quantitatively differ from the case of isotropic interaction. The crucial difference occurs in the flow of $ J_{xy} $.
Inserting the modified gap equation~(\ref{eqn:xxzdelta}) into Eq.~(\ref{eqn:xxz23}) yields
\begin{equation} \label{eqn:mod-j23}
 J_{xy} = \frac{4 J \left( 1 + \alpha \right) \Delta}{\Delta_0 + \alpha \Delta} \, ,
\end{equation}
which gives finite values in the limit $ \lambda, \Delta_0 \to 0 $.

\begin{figure}
 \includegraphics[width=8.5cm]{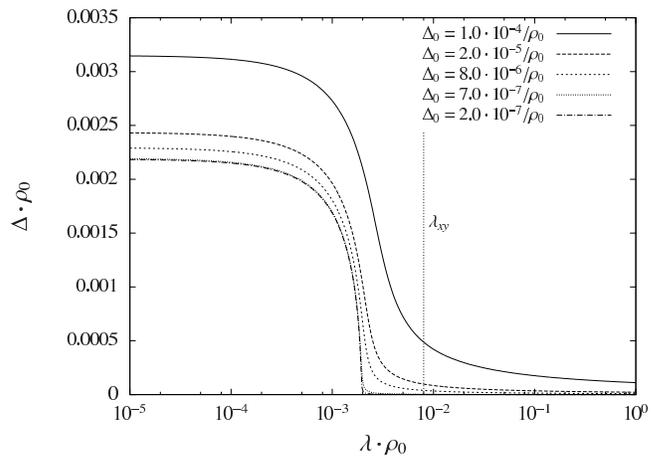}
 \caption{Flow of $\Delta $ according to Eqs.~(\ref{eqn:flow-13})  and (\ref{eqn:flow-delta}) for $ \alpha =- 0.2$, zero temperature, perfect nesting and a constant density of states $\rho_0$
 (all other parameters as in Fig.~\ref{fig:delta-alpha}). Below $ \lambda_{xy} \approx 8 \cdot 10^{-3} / \rho_0 $, the results are unphysical.} \label{fig:delta-alpha-}
\end{figure}
\begin{figure}
 \includegraphics[width=8.5cm]{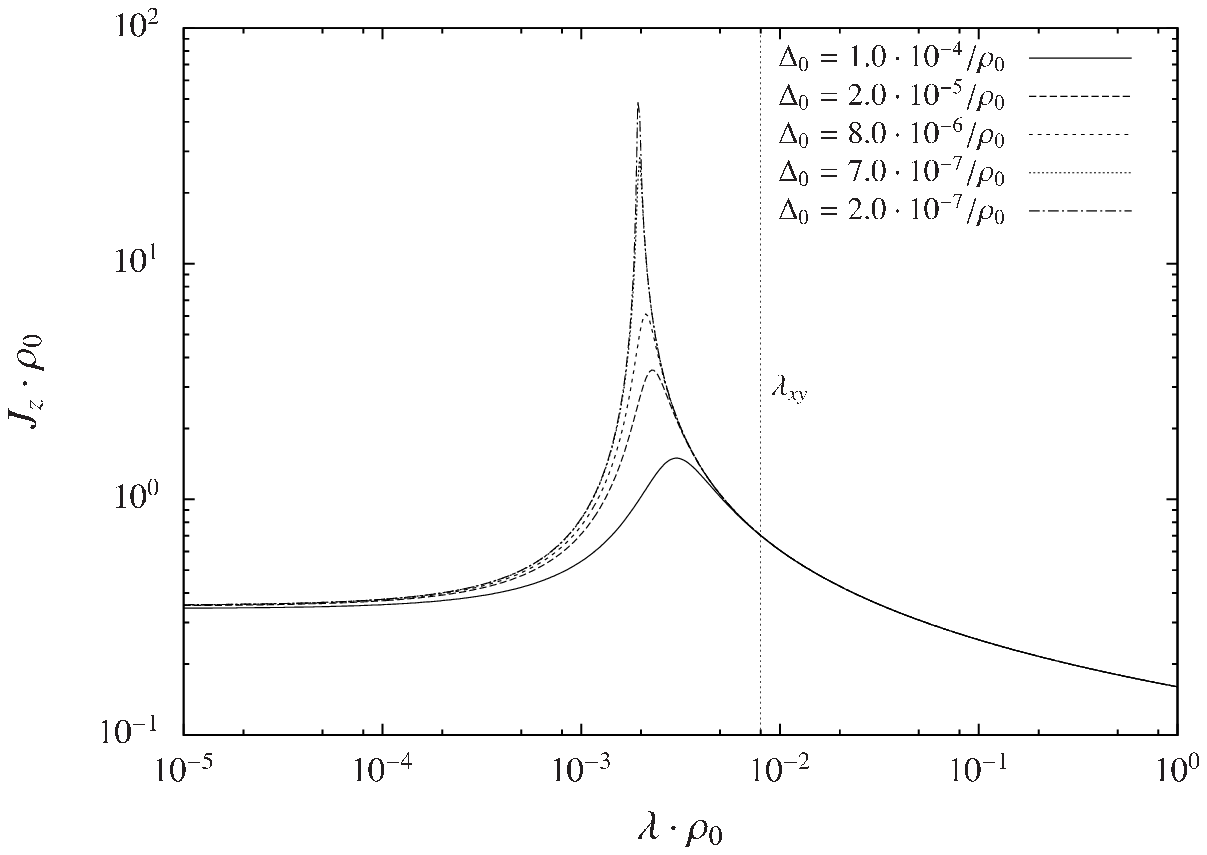}
 \caption{Flow of $J_z $ according to Eqs.~(\ref{eqn:flow-13}) and (\ref{eqn:flow-delta}) for $ \alpha =- 0.2$, zero temperature, perfect nesting, and a constant density of states $\rho_0$
 (all other parameters as in Fig.~\ref{fig:delta-alpha}). Below $ \lambda_{xy} \approx 8 \cdot 10^{-3} / \rho_0 $, the results are unphysical.} \label{fig:j13-alpha-}
\end{figure}
\begin{figure}
 \includegraphics[width=8.5cm]{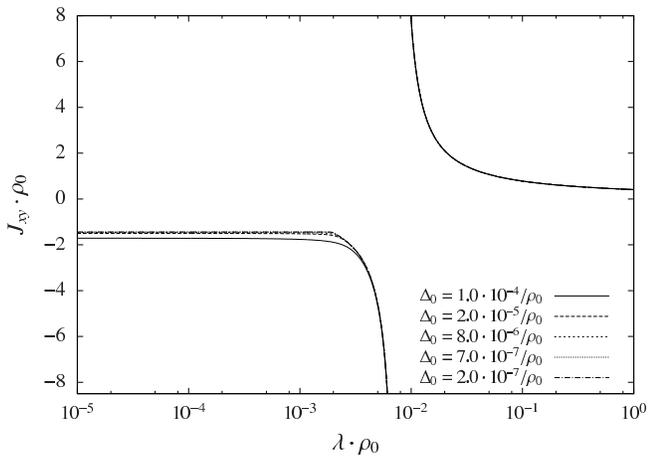}
 \caption{Flow of $J_{xy} $ according to Eq.~(\ref{eqn:mod-j23}) for $ \alpha =- 0.2$, zero temperature, perfect nesting, and a constant density of states $\rho_0$
 (all other parameters as in Fig.~\ref{fig:delta-alpha}). At scales below the sign change of $ J_{xy} $, the results are unphysical.} \label{fig:j23-alpha-}
\end{figure}
We now consider the flow in the $xy$-regime with an anisotropy factor $ \alpha = -0.2 $ (see Figs.~\ref{fig:delta-alpha-}--\ref{fig:j23-alpha-}). Therefore the spins should be aligned in the $xy$-plane rather than in the $ z$-direction in the ground state. Since our formalism does not allow for a staggered in-plane magnetization, we should encounter an instability corresponding to spins ordered along a direction in the $ xy $-plane. The flow of the gap and $ J_z $ does not seem to differ much from the previous case.
From Eq.~(\ref{eqn:mod-j23}), we however find that for $ -1 < \alpha < 0 $, $ J_{xy} $ undergoes a sign change at some scale $ \lambda_{xy} $ as shown in Fig.~\ref{fig:j23-alpha-}. Of course, this divergence forbids direct integration of the flow equation~(\ref{eqn:flow-23}) down to the infrared. Instead, a flow to strong coupling is observed. According to Fig.~\ref{fig:j23-alpha-}, this instability occurs at about $ \lambda_{xy} = 8 \cdot 10^{-3} /\rho_0 $. At this scale, both $ \Delta $ and $ J_z$ remain finite.
Formally, the flow of $ J_z $ and $ \Delta $ (Eqs.~(\ref{eqn:flow-13})  and (\ref{eqn:flow-delta})) can be continued below $\lambda_{xy} $. The result obtained is similar to the previous cases (see Figs.~\ref{fig:delta-alpha-} and \ref{fig:j13-alpha-}). Again $ J_z $ has peak at a scale $ \lambda_z < \lambda_{xy} $, where $\Delta$ starts to grow significantly.
 Note however that the formal solution of the flow equations~(\ref{eqn:flow-13}) for $J_z$ and (\ref{eqn:flow-delta}) for $\Delta$ becomes unphysical at scales $ \lambda < \lambda_{xy} $, as the instability corresponds to a staggered magnetization in the $ xy $-plane. Our seed field, however, was set to zero in the $ x $ and $ y$ directions \emph{before} the thermodynamic limit was taken, which is not the physical order of the limits for spontaneous symmetry breaking. Of course, with some additional effort, we could now allow for a $xy$-symmetry-breaking component of the self-energy in straight analogy to what we did so far for the $z$ direction. That way, one could follow the flow further through the $xy$-instability. At the level presented here, we can infer that the formalism with symmetry breaking only in the $z$ direction is faithful enough to allow for additional instabilities. This is important in order to preserve the unbiased character of the RG approach. 
\section{Channel decomposition} \label{sec:ch-dec}
So far, we only have considered the RG flow for mean-field models. In realistic models showing an antiferromagnetic instability of itinerant electrons, however, transfer momenta other than the ordering vector $ {\bf Q} $ play a role in the particle-hole channel, which, in addition, couples with the particle-particle channel. If one studies the behavior of the effective interaction for wave-vector transfers near ${\bf Q}$, one will find that there is a sharp peak at ${\bf Q}$. Similar statements hold for the frequency dependence, where a sharp peak forms around zero transfer frequency. As one will have to resort to a numerical evaluation of the flow equations and the corresponding wave vector and frequency summations in most cases, such sharp peaks pose a challenge to the usual discretization schemes. In the general case, these sharp structures can occur for both of the two possible  wave-vector/frequency transfers in the interaction $t=k_1-k_3$ and $u=k_2-k_3$, and for the total wave vector/frequency $s=k_1+k_2$. Hence, the interaction as function of $k_1$, $k_2$, and $k_3$ can possibly depend strongly on all these three variables.
Here, a vertex decomposition proposed by Husemann and Salmhofer\cite{husemann_09,husemann_12} for the wave-vector dependence and by Karrasch {\sl et al}.\cite{karrasch} for the frequency dependence can  serve as a simplifying approximation. The idea is to write the interaction as a sum of (at least) three functions. One of these functions is used to capture the strong $s$ dependence of the interaction, one is used for the $t$ dependence and one for the $u$ dependence, as will be detailed below. Then, instead of having one function that depends strongly on three variables, one has three functions that depend strongly only on one variable. Hence, the discretization effort can be reduced from $N^3$ down to $3N$. The idea is based on the observation that the second-order corrections for constant bare interactions fulfill this decomposition exactly. One may hope that higher orders are still approximated well.

So far, this decomposition has been proposed and tested for the symmetric phase, which gave sufficient evidence for the validity of the underlying approximation. Recently, an extension to the superfluid phase has been given by Eberlein and Metzner.\cite{eberlein_param,eberlein-unpub} In this section, we propose a channel decomposition in the phase without spin-rotational symmetry, which should make a numerical solution of the resulting flow equations tractable also in this case.

For commensurate AFM, the renormalized interaction is only invariant under translations by an even number of sites. In momentum space, this means that the coupling functions are of the form
\begin{align*}
 V_X & (k_1,k_2,k_3,k_4) = V_X^{\rm c} (k_1,k_2,k_3) \, \delta \left( k_1 + k_2 - k_3 - k_4 \right) \\
& +V_X^{\rm nc} (k_1,k_2,k_3) \,\, \delta \left( k_1 + k_2 - k_3 - k_4  +Q \right) \, ,
\end{align*}
with a momentum-conserving  part $ V_X^{\rm c} $ and a non-conserving part $ V_X^{\rm nc} $, which is generated during the flow. In Nambu representation with spinors according to Eq.~(\ref{eqn:Nambu-spinor}),
the interaction can be parameterized in the same way with coupling functions $ W_X (K_1,K_2,K_3,K_4) $ as in the conventional representation with coupling functions $ V_X (k_1,k_2,k_3,k_4)$, where $ K_i = (k_i,s_i) $ with Nambu indices $ s_i$.
The components $ W_X^{\left\{s_i\right\}} $ of the interaction in Nambu representation with an even number of equal Nambu indices $ s_i = \pm 1 $ then correspond to $ V_X^{\rm c}$ while the components with an odd number of equal Nambu indices correspond to $ V_X^{\rm nc}$. Thus, we have
\begin{align*}
 W_X (K_1,K_2,K_3,K_4) & = \tilde{\delta}_{\left\{ k_i \right\}} \\ 
  \times  & \left\{ \begin{array}{ll}  V_X^{\rm c} (\varkappa_1,\varkappa_2,\varkappa_3) & \sum_i \frac{s_i}{2} \,\, \text{even} \\ V_X^{\rm nc} (\varkappa_1,\varkappa_2,\varkappa_3) & \sum_i \frac{s_i}{2} \,\, \text{odd} \end{array} \right. \, ,
\end{align*}
with physical momenta $ \varkappa_i = k_i + \left( 1 - s_i \right) Q /2 $.
In this formula, the momenta $ {\rm k}_i $ are restricted to half the Brillouin zone, and therefore $ \tilde{\delta}_{\left\{ k_i \right\} } $ ensures momentum conservation up to multiples of $ \pi $.
We now decompose the three coupling functions from Sec.~\ref{sec:param-gen} as follows.
Renormalizations of  equal-spin interactions $ W_\uparrow $ and $ W_\downarrow $ can be regarded as a sum of triplet-superconductivity contributions and a spin-dependent CDW term:
\begin{align*}
 W_\tau (K_1,K_2,K_3,K_4)  = & \, \tilde{\delta}_{\left\{ k_i \right\} } \left[ U_\tau^{\left\{ s_i \right\} }  (k_1,k_2,k_3) \right. \\ &+ \Phi_{{\rm tSC},\tau}^{\left\{ s_i \right\} }  (k_1+k_2,k_1,k_3)\\ & + \Phi_{{\rm CDW},\tau}^{\left\{ s_i \right\} }  (k_3-k_2,k_1,k_2) \\ & - \left. \Phi_{{\rm CDW},\tau}^{\left\{ \tilde{s}_i \right\} }  (k_1-k_3,k_1,k_2) \right] \, , 
\end{align*}
where $ U_\tau $ stems from the bare interaction and where $ \tilde{\bf s} =(s_1,s_2,s_4,s_3) $. The coupling function $ W_{\uparrow \downarrow} $, in contrast, is renormalized by a particle-particle coupling-function, which may contain triplet- as well as singlet-superconductivity terms, and magnetic contributions corresponding to $ S_x^2 + S_y^2 $ and $ S_z^2 $ terms in the interaction
\begin{align*}
 W_{\uparrow \downarrow} (K_1,K_2,K_3,K_4)  =  & \, \tilde{\delta}_{\left\{ k_i \right\} } \left[ U_{\uparrow \downarrow}^{\left\{ s_i \right\} }  (k_1,k_2,k_3) \right. \\ & + \Phi_{{\rm SC}\uparrow,\downarrow}^{\left\{ s_i \right\} }  (k_1+k_2,k_1,k_3)\\ &+ \Phi_{{\rm M},xy}^{\left\{ s_i \right\} }  (k_3-k_2,k_1,k_2) \\ & + \left. \Phi_{{\rm M},z}^{\left\{ s_i \right\} }  (k_1-k_3,k_1,k_2)\right] \, . 
\end{align*}
The right-hand sides of the flow equations (\ref{eqn:start-ffe})--(\ref{eqn:stop-ffe}) now read as follows.
We use 
\begin{equation*}
 L_{\tau_1,\tau_2}^{\left\{ s_i \right\} } (p,q) = \partial_\lambda \left[ G_{\tau_1}^{s_1,s_2} (p) \,G_{\tau_2}^{s_3,s_4} (q) \right]
\end{equation*}
as a short-hand notation for the loops, and the measure $ d'p $ indicates that the respective momentum integral only runs over the folded Brillouin zone.
In the particle-particle channel, one obtains
\begin{widetext}
\begin{align*}
 \dot{\Phi}_{{\rm tSC}, \tau}^{\left\{ s_i \right\}} (l,q,q') & = \frac{1}{2} \sum_{\left\{ s'_i \right\}} \int \!\! d' p \, W_\tau^{s_1,s_2,s'_1,s'_3} (q,l-q,p,l-p) \, W_\tau^{s'_4,s'_2,s_3,s_4} (l-p,p,q',l-q') \, L_{\tau,\tau}^{\left\{ s'_i \right\} } (p,l-p)\, , \\
 \dot{\Phi}_{{\rm SC}, \uparrow \downarrow}^{\left\{ s_i \right\}} (l,q,q') & = - \sum_{\left\{ s'_i \right\}} \int \!\! d' p \, W_{\uparrow \downarrow}^{s_1,s_2,s'_1,s'_3} (q,l-q,p,l-p) \, W_{\uparrow \downarrow}^{s'_2,s'_4,s_3,s_4} (p,l-p,q',l-q') \, L_{\uparrow,\downarrow}^{\left\{ s'_i \right\} } (p,l-p)  \, .
\end{align*}
The particle-hole channel gives
\begin{align*}
  \dot{\Phi}_{{\rm CDW}, \uparrow}^{\left\{ s_i \right\}} (l,q,q') & = -  \sum_{\left\{ s'_i \right\}} \int \!\! d' p \, W_\uparrow^{s'_4,s_2,s_3,s'_1} (l+p,q',l+q',p) \, W_\uparrow^{s_1,s'_2,s'_3,s_4} (q,p,l+p,q-l)  \,L_{\uparrow,\uparrow}^{\left\{ s'_i \right\} } (p,l+p)  \\
& \quad  -  \sum_{\left\{ s'_i \right\}} \int \!\! d' p \, W_{\uparrow \downarrow}^{s_2,s'_4,s_3,s'_1} (q',l+p,l+q',p) \, W_{\uparrow \downarrow}^{s_1,s'_2,s_4,s'_3} (q,p,q-l,l+p)  \,L_{\downarrow,\downarrow}^{\left\{ s'_i \right\} } (p,l+p)  \, ,
\end{align*}
\begin{align*}
  \dot{\Phi}_{{\rm CDW}, \downarrow}^{\left\{ s_i \right\}} (l,q,q') & = -  \sum_{\left\{ s'_i \right\}} \int \!\! d' p \, W_\downarrow^{s'_4,s_2,s_3,s'_1} (l+p,q',l+q',p) \, W_\downarrow^{s_1,s'_2,s'_3,s_4} (q,p,l+p,q-l)  \,L_{\downarrow,\downarrow}^{\left\{ s'_i \right\} } (p,l+p)  \\
& \quad  -  \sum_{\left\{ s'_i \right\}} \int \!\! d' p \, W_{\uparrow \downarrow}^{s'_4,s_2,s'_1,s_3} (l+p,q',p,l+q') \, W_{\uparrow \downarrow}^{s'_2,s_1,s'_3,s_4} (p,q,l+p,q-l)  \,L_{\uparrow,\uparrow}^{\left\{ s'_i \right\} } (p,l+p)  \, ,
\end{align*}
\begin{align*}
  \dot{\Phi}_{{\rm M}, xy}^{\left\{ s_i \right\}} (l,q,q') & = -  \sum_{\left\{ s'_i \right\}} \int \!\! d' p \, W_{\uparrow \downarrow}^{s'_4,s_2,s_3,s'_1} (l+p,q',l+q',p) \, W_{\uparrow \downarrow}^{s_1,s'_2,s'_3,s_4} (q,p,l+p,q-l) \, L_{\downarrow,\uparrow}^{\left\{ s'_i \right\} } (p,l+p)  \, ,
\end{align*}
\begin{align*}
  \dot{\Phi}_{{\rm M}, z}^{\left\{ s_i \right\}} (l,q,q') & = -  \sum_{\left\{ s'_i \right\}} \int \!\! d' p \, W_{\uparrow \downarrow}^{s'_4,s_2,s'_1,s_4} (l+p,q',p,q'+l) \, W_\uparrow^{s_1,s'_2,s'_3,s_3} (q,p,l+p,q-l) \, L_{\uparrow,\uparrow}^{\left\{ s'_i \right\} } (p,l+p) \\
 & \quad -  \sum_{\left\{ s'_i \right\}} \int \!\! d' p \, W_\downarrow^{s'_4,s_2,s_4,s'_1} (l+p,q',l+q',p) \, W_{\uparrow \downarrow}^{s_1,s'_2,s_3,s'_3} (q,p,q-l,l+p) \, L_{\downarrow,\downarrow}^{\left\{ s'_i \right\} } (p,l+p)  \, .
\end{align*}
\end{widetext}
This channel decomposition should allow for efficient further parametrization giving rise to numerically tractable flow equations. Nevertheless, implementing and testing this scheme represents a longer effort, and at this point we can not present any specific results.
\section{Summary and outlook} \label{sec:conc}
In this work, we have parameterized the effective action and given fRG flow-equations for spin-$1/2$ fermions in phases with broken spin invariance. One then has to deal with three coupling functions for the interactions instead of one in the symmetric phase. The flow of the AFM mean-field model can be captured by three running couplings corresponding to radial and Goldstone components of the interaction and the gap. As one may have expected, these couplings behave as in previous studies\cite{brokensy_SHML,gersch_cdw} of other mean-field models. The radial coupling $ J_z $ shows a pronounced hump at the critical scale, where the gap $ \Delta $ starts to grow significantly. While the radial coupling and the gap saturate to finite infrared values for small seed fields, the Goldstone coupling $ J_{xy} $ diverges, reflecting the masslessness of the Goldstone modes. This can as well be seen from the global SU(2) Ward identity. The RG flow equations reproduce the RPA result for the interaction and the self-consistent gap equation exactly. This suggests that we have indeed found a sensible parametrization. A generalization to a symmetry-breaking interaction of $ xxz$ type is straightforward. In the Ising regime, the Goldstone modes are then gapped, which renders all couplings finite in the infrared. For negative anisotropy parameters of the bare interaction, i.e., in the $ xy $ regime, $ J_{xy} $ flows to strong coupling. This instability corresponds to the ordering of spins in the $ xy $ plane. However, this type of spontaneous symmetry breaking is not included in the formalism presented here (but could be included with some additional work). This example shows that the built-in symmetry breaking in the $z$ direction going along with a gapping of the Fermi surface does not prevent the physically reasonable instability toward ordering in the $xy$ plane. Hence, at least on this level, triggering symmetry breaking in a specific direction does not remove the usual RG advantage of providing an unbiased description.

For an isotropic bare interaction, this formalism can also be used for proceeding beyond the level of a reduced mean-field model. Then, the usual advantage of RG approaches, which is to include all the different one-loop diagrams, comes into play and can lead to significant deviations from the mean-field picture. However, on the technical level various complications arise: particle-particle diagrams lead to a non-trivial momentum and frequency dependence of the coupling functions. Since this has to be resolved numerically or by a suitable ansatz, we expect that this drastically increases the computational cost. Hence, a channel decomposition\cite{husemann_09,husemann_12} appears to be in order. This decomposition replaces an interaction that may depend strongly on three wave vectors and frequencies by (at least) three functions that depend strongly only on one wave-vector sum or transfer. This reduces the numerical effort considerably.
For commensurate AFM, the symmetry-breaking one-particle term violates momentum conservation and non-conserving contributions to the interaction are generated in the flow. In the commensurate case, the Nambu formalism helps to facilitate our notation. We therefore have derived channel-decomposed flow equations in Nambu representation. The next step will indeed be the implementation of this scheme for non-mean-field models. As this is still  a larger numerical task, we refrain from embarking on this here and only describe the theoretical foundation. 
In solving the flow equations for the more general case, one may proceed as for a fermionic superfluid.\cite{eberlein-unpub} However, especially in the case of AFM, an additional feature may be of interest. In Ref.~\onlinecite{eberlein-unpub}, the momentum dependence of both the fermion-boson vertices and the gap was assumed to correspond to a simple angle-independent $ s $-wave, which is a reasonable approximation for the attractive Hubbard model. (However, the approach pursued in that work can in principle account for more complicated boson-fermion vertices.)  In the repulsive Hubbard model at small but non-zero second-neighbor hopping the situation is different in two ways. For one thing, the particle-particle contributions to the interaction may then develop some structure that has an impact on the fermion-boson vertices in the particle-hole channels, e.g., on their magnitude and dependence on the fermionic wave vectors. It may therefore be advantageous to track
the deformation of fermionic form factors in the flow. Furthermore, a reconstructed Fermi surface might still give rise to a Cooper instability \emph{inside} the AFM symmetry-broken phase, which would account for a homogeneous coexistence of non-zero AFM and superconductivity order parameters. In principle, this effect should be captured in our formalism, but its demonstration remains an issue of future research.

\section*{Acknowledgments}
We thank A.~Eberlein, A.~A.~Katanin, W.~Metzner, and M.~Salmhofer for useful discussions. This project was supported by the German Research Foundation DFG via FOR 723. 
\bibliography{biblio}
\end{document}